\documentclass[11pt]{article} 

\usepackage{amsmath}
\usepackage{amssymb}

\usepackage{graphicx}

\usepackage{color}

\usepackage[table]{xcolor} 
\rowcolors{2}{white}{gray!25}

\usepackage{fancyhdr}

\usepackage[labelfont=bf, small]{caption}
\usepackage[numbers]{natbib}
\newcommand{\argmin}{\operatorname{argmin}\displaylimits}

\long\def\symbolfootnote[#1]#2{\begingroup%
\def\thefootnote{\fnsymbol{footnote}}\footnote[#1]{#2}\endgroup}

\date{}

\begin{document}

\begin{center}
{\Large
\textbf{Boosting the concordance index for survival data -- a unified framework to derive and evaluate biomarker combinations}\symbolfootnote[1]{
\textit{This is a preliminary version of an article submitted to PLOS ONE. If citing, please refer to the original article.}
 }} \vspace{0.75cm}
\\
Andreas Mayr$^{1}$,
Matthias Schmid$^{2}$ \vspace{0.5cm}
\\
$^{1}$Department of Medical Informatics, Biometry and
Epidemiology, Friedrich-Alexander-Universit\"at Erlangen-N\"urnberg, Erlangen, Germany
\\
$^{2}$Department of Statistics, Ludwig-Maximilians-Universit\"at M\"unchen, Munich, Germany
\end{center}
\setlength{\parindent}{0cm} 
\vspace{0.5cm}

\begin{abstract}
The development of molecular signatures for the prediction of time-to-event outcomes is a methodologically challenging task in bioinformatics and biostatistics. Although there are numerous approaches for the derivation of marker combinations and their evaluation, the underlying methodology often suffers from the problem that different optimization criteria are mixed during the feature selection, estimation and evaluation steps. This might result in marker combinations that are suboptimal regarding the evaluation criterion of interest. To address this issue, we propose a unified framework to derive and evaluate biomarker combinations. Our approach is based on the concordance index for time-to-event data, which is a non-parametric measure to quantify the discriminatory power of a prediction rule. Specifically, we propose a gradient boosting algorithm that results in linear biomarker combinations that are optimal with respect to a smoothed version of the concordance index. We investigate the performance of our algorithm in a large-scale simulation study and in two molecular data sets for the prediction of survival in breast cancer patients. Our numerical results show that the new approach is not only methodologically sound but can also lead to a higher
discriminatory power than traditional approaches for the derivation of gene signatures.
\end{abstract}

\section*{Introduction}

Recent technological developments in the fields of genomics and biomedical research have led to the discovery of large numbers of gene signatures for the prediction of clinical survival outcomes. In cancer research, for example, gene expression signatures are nowadays used to predict the time to occurrence of metastases \cite{desmedt, vandevijver} as well as the time to progression \cite{kok} and overall patient survival \cite{ligui, chang}. While the importance of molecular data in clinical and epidemiological research is expected to grow considerably in the next years \cite{microrna_plosone, personalized_nature, witten2010survival}, the detection of clinically useful gene signatures remains a challenging problem for bioinformaticians and biostatisticians, especially when the outcome is a survival time.

After normalization and data pre-processing, the development of a new gene signature usually comprises three methodological tasks:
\begin{enumerate}
\item[{}] Task 1: Select a subset of genes that is associated with the clinical outcome.
\item[{}] Task 2: Derive a marker signature by finding the ``optimal'' combination of the selected genes.
\item[{}] Task 3: Evaluate the prediction accuracy of the optimal combination using future or external data.
\end{enumerate}

Task 1, the selection of a clinically relevant subset of genes, is often addressed by calculating scores to rank the univariate association between the survival outcome and each of the genes \cite{masong, witten2010survival}. In a subsequent step, the genes with the strongest associations are selected to be included in the gene signature.

Task 2, the derivation of an optimal combination of the selected genes, is usually fulfilled by forming linear combinations of gene expression levels based on Cox regression. Due to multicollinearity problems and the high dimensionality of molecular data, a direct optimization of the Cox partial likelihood is often unfeasible \cite{witten2010survival}. Consequently, marker combinations are often derived by combining coefficients of univariate Cox regression models \cite{wanglancet}, or by applying regularized Cox regression techniques (such as the Lasso \cite{tibshCoxlasso, goemanlasso} or ridge-penalized regression \cite{li2002Ridge, gui2005penalized}).

Task 3, the evaluation of  prediction accuracy, is considered to be a challenging problem in survival analysis. This is because traditional performance measures for continuous outcomes (such as the mean squared error) are no longer applicable in the presence of censoring. In the literature, several approaches to address this problem exist (see, e.g., \cite{schmidpotapov} for an overview). In this article, we focus on the {\it concordance index for time-to-event data} ($C$-index \cite{harrell1, harrell2, unoC}), which has become a widely used measure of the performance of biomarkers in survival studies \cite{caspar, pzj,zhang_plosone, zhao_plosone}. Briefly spoken, the $C$-index can be interpreted as the probability that a patient with a small survival time is associated with a high value of a biomarker combination (and vice versa). Consequently, it measures the concordance between the rankings of the survival times and the biomarker values and therefore the ability of a biomarker to discriminate between patients with small survival times and patients with large survival times. This strategy is especially helpful if the aim is to subdivide patients into groups with good or poor prognosis (as applied in many articles in the medical literature, e.g., \cite{wanglancet}). By definition, the $C$-index has the same scale as the classical area under the curve (AUC) in binary classification: While prediction rules based on random guessing yield $C = 0.5$, a perfectly discriminating biomarker combination leads to $C = 1$.

Interestingly, the derivation of new gene signatures for survival outcomes via Tasks 1--3 is often addressed by combining completely different methodological approaches and estimation techniques. For example, the estimation of biomarker combinations is usually based on Cox regression and is hence carried out via the optimization of a partial likelihood criterion. On the other hand, the resulting combinations are often evaluated by using the $C$-index \cite{zhang_plosone, zhao_plosone} which has its roots in the receiver operating characteristics (ROC) methodology. This methodological inconsistency is also problematic from a practical point of view, as the marker combination that optimizes the partial log likelihood criterion is not necessarily the one that optimizes the $C$-index. In other words, if the $C$-index and therefore the discriminatory power is the evaluation criterion of interest, it may be suboptimal to use a likelihood-based criterion to optimize the marker combination. This issue is, of course, not only problematic in survival analysis but also in regression and classification. A theoretical discussion on the differences between performance measures for binary classification can, e.g., be found in \cite{friedman_bias}.

To overcome the aforementioned inconsistencies, we propose a unified framework for survival analysis that is based on the same statistical methodology for gene selection (Task 1), derivation of an optimal biomarker combination (Task 2) {\it and} the evaluation of a new gene signature (Task 3). As will be demonstrated, all three tasks can be addressed by using the concordance index for time-to-event data as performance criterion. While the $C$-index has already been proposed for gene selection (Task 1) and the evaluation of prediction accuracy (Task 3) \cite{masong, schmidpotapov}, the main contribution of this article is a new estimation technique that addresses the development of optimal combinations of genes (Task 2). To achieve this goal, we propose a method for finding gene combinations that is based on the gradient boosting framework \cite{BuhlmannHothorn06}. As will be shown, it is possible to use boosting to derive prediction-optimized gene combinations via direct optimization of the $C$-index. Because this new approach allows for using the $C$-index to address all three tasks, the proposed method leads to a consistent framework for the derivation of gene signatures in biomarker studies where the $C$-index is the main performance criterion.


\section*{Methods}

\subsection*{Notation}

We first introduce some basic notation that will be used throughout the article. Let $T\in\mathbb{R}^+$ be the survival time of interest and
$X = (x_1, ..., x_p) \in\mathbb{R}^p$ a vector of predictor variables. In addition to the gene expression levels, $X$ may contain the measurements of some clinical predictor variables. Denote the conditional survival function of $T$ given $X$ by $S(t|x)
= \mathrm{P}(T>t |X=x)$. The probability density and survival functions of $T$ are denoted by $f(t)$ and $S(t)$, respectively. Further let $T_{\text{cens}}\in\mathbb{R}^+$ be a random censoring time and denote the observed survival time by $\tilde{T} := \min (T,T_{\text{cens}})$. The random variable $\Delta := \mathrm{I} (T \le T_{\text{cens}})$ indicates whether $\tilde{T}$ is right-censored ($\Delta=0$) or not ($\Delta=1$).

A prediction rule for $T$ will be formed as a linear combination
\begin{equation}
\eta  := \beta_0 + \sum_{l =1}^p \beta_l \cdot x_l  = X^{\top}{\boldsymbol \beta} \, ,
\end{equation}
where ${\boldsymbol \beta}$ is an unknown vector of coefficients.
We generally assume that the estimation of $\hat{\beta}$ is based on an i.i.d. learning sample $\{ (\tilde{T}_i^L, \Delta_i^L, X_i^L),\, i=1,\dots,n\}$. In case of the Cox regression model, for example, $\eta$ is related to $T$ by the equation
\begin{equation}
S(t|x) = \exp \big( - \Lambda_0 (t) \cdot \exp \big( \eta \big) \big) \, ,
\end{equation}
where $\Lambda_0 (t)$ is the cumulative baseline hazard function. Because there is a one-to-one relationship between $\eta$ and the expected survival
time $\mathrm{E}(T|X)$, the linear combination $\eta$ can be used as a biomarker to predict the survival of individual patients.

\subsection*{Concordance index}

Our proposed framework to derive and evaluate biomarker combinations is based on the {\em concordance index} (``$C$-index'') which is a general discrimination measure for the evaluation of prediction models \cite{harrell1, harrell2}. It can be applied to continuous, ordinal and dichotomous outcomes \cite{harrell3}. For time-to-event outcomes, the $C$-index is defined as
\begin{equation} \label{CindexProb}
C := \mathrm{P} (\eta_{j_1} > \eta_{j_2} \, | \, T_{j_1} < T_{j_2}) \, ,
\end{equation}
where $T_{j_1}$, $T_{j_2}$ and $\eta_{j_1}$, $\eta_{j_2}$ are the event times and the predicted marker values, respectively, of two observations in an i.i.d. test sample $\{(\tilde{T}_j,\Delta_j,X_j), j=1,\dots,N \}$. By definition, the $C$-index for survival data measures whether large values of $\eta$ are associated with short survival times $T$ and vice versa. Moreover, it can be shown that the $C$-index is equivalent to the area under the time-dependent ROC curve,  which is a measure of the discriminative ability of $\eta$ at each time point under consideration (see \cite{heagerty05}, p. 95 for a formal proof).

During the last decades, the $C$-index has gained enormous popularity in biomedical research; for example, searching for the terms ``concordance index'' and ``c-index'' in PubMed \cite{PubMed} resulted in 1156 articles by the time of writing this article. Generally, a value of $C$ close to 1 indicates that the marker $\eta$ is close to a perfect discriminatory power, while a marker that does not perform better than chance results in a value of 0.5. For example, the famous Gail model \cite{gailmodel} for the prediction of breast cancer is estimated to yield a value of $C=0.67$ \cite{gail_c}.

Being a flexible discrimination measure, the $C$-index is especially useful for selecting and ranking genes from a pre-processed set of high-dimensional gene expression data (Task 1 described in the Introduction). In other words, Task 1 can be addressed by computing the $C$-index (and hence the marginal discriminatory power) for each individual gene or biomarker, where only those genes with the highest $C$-index are incorporated into the yet-to-derive optimal combination (Task 2). Although there exist various other ways to rank genes and select the most influential ones, the $C$-index has been demonstrated to be especially advantageous for this task \cite{masong}.

An estimator of the $C$-index for survival data is given by

\begin{equation}
\label{harrellsc} \widehat{C}_{\mathrm{surv}} \ := \ \frac{\sum_{j < k}
\mathrm{I} \left( \tilde{T}_j < \tilde{T}_k \right) \mathrm{I}
\left( \hat{\eta}_j > \hat{\eta}_k \right) \Delta_j + \mathrm{I}
\left( \tilde{T}_k < \tilde{T}_j \right) \mathrm{I} \left(
\hat{\eta}_k > \hat{\eta}_j \right) \Delta_k}{\sum_{j < k}
\mathrm{I} \left( \tilde{T}_j < \tilde{T}_k \right) \Delta_j +
\mathrm{I} \left( \tilde{T}_k < \tilde{T}_j \right) \Delta_k }
\end{equation}

with $j,k \in \{1, \dots, n\}$ (``Harrell's $C$ for survival data'', \cite{antolini}). Generally, $\widehat{C}_{\mathrm{surv}}$ is a consistent estimator of the $C$-index in situations where no censoring is present. However, because pairs of observations where the smaller observed survival time is censored are ignored in formula (\ref{harrellsc}), $\widehat{C}_{\mathrm{surv}}$ is known to show a notable upward bias in the presence of censoring. This bias tends to be correlated with the censoring rate \cite{antolini, schmidpotapov}.

To overcome the censoring bias of $\widehat{C}_{\mathrm{surv}}$, Uno et al. \cite{unoC} proposed a modified version of $\widehat{C}_{\mathrm{surv}}$, which is defined as

\begin{equation}
\label{unoc} \widehat{C}_{\mathrm{Uno}} \ := \ \frac{\sum_{j,k} (
\hat{G}_{n}^{L} ( \tilde{T}_j ) )^{-2} \, \mathrm{I}
\left( \tilde{T}_j < \tilde{T}_k \right) \mathrm{I} \left(
\hat{\eta}_j > \hat{\eta}_k \right) \Delta_j}{\sum_{j,k} (
\hat{G}_{n}^{L} ( \tilde{T}_j ) )^{-2} \, \mathrm{I}
\left( \tilde{T}_j < \tilde{T}_k \right) \Delta_j } \ ,
\end{equation}

where $\hat{G}_{n}^{L}(t)$ denotes the Kaplan-Meier
estimator of the unconditional survival function of $T_\text{cens}$ (estimated from the learning data). In the following, we will assume that the censoring times are independent of $X$. Under this assumption,  $\widehat{C}_{\mathrm{Uno}}$ is a consistent and asymptotically normal estimator of $C$ (see \cite{unoC}, pp. 1113--1115). Consistency  is ensured by applying inverse probability weighting (using the weights $\Delta_j / (\hat{G}_{n}^{L}( \tilde{T}_j)^{2})$, which account for the inverse probability that an observation in the test data is censored \cite{vanderlaan}). Numerical studies suggest that $\widehat{C}_{\mathrm{Uno}}$ is remarkably robust against violations of the random censoring assumption \cite{schmidetalBriefings}.

Apart from the estimators $\widehat{C}_{\mathrm{surv}}$ and $\widehat{C}_{\mathrm{Uno}}$, there exist various other approaches to estimate the probability in (\ref{CindexProb}) (see, e.g., \cite{schmidpotapov} for an overview). Most of these approaches are based on the assumptions of a Cox proportional hazards model, so that they are not valid in case these assumptions are violated. Because $\widehat{C}_{\mathrm{Uno}}$ is model-free and because the consistency of $\widehat{C}_{\mathrm{Uno}}$ is guaranteed even in situations where censoring rates are high (in contrast to the estimator $\widehat{C}_{\mathrm{surv}}$), we will base our boosting method on $\widehat{C}_{\mathrm{Uno}}$.

\subsection*{Boosting the concordance index}

The core of our proposed framework to address Tasks 1 -- 3 is the derivation of a prediction-optimized linear combination of genes that is optimal w.r.t. to the $C$-index for time-to-event data. Our approach will be based on a component-wise gradient boosting algorithm \cite{BuhlmannHothorn06} that uses the $C$-index as optimization criterion.

Gradient boosting algorithms \cite{friedmanetal2000} are generally based on a loss function $\rho (T, \eta)$ that is assumed to be differentiable with respect to the predictor $\eta \equiv \eta (X)$. The aim is then to estimate the ``optimal'' prediction function
\begin{equation}
\label{expmean}  \eta^{*} := \argmin_{\eta} \mathbb{E}_{T,X} \left[\rho (T, \eta(X)) \right]
\end{equation}
by using gradient descent techniques. As the theoretical mean in (\ref{expmean}) is usually unknown in practice, gradient boosting algorithms minimize the empirical risk $\mathcal{R} := \sum_{i=1}^n \rho (t_i,\eta(x_i))$ over $\eta$ instead.

When considering the $C$-index for survival data, the aim is to determine the optimal predictor~$\eta^*$ that maximizes the concordance measure $C = \mathrm{P} (\eta^*_{i} > \eta^*_{k} \, | \, T_{i} < T_{k})$ -- which is essentially the solution to Task 2. Hence a natural choice for the empirical risk function $\mathcal{R}$ would be the negative concordance index estimator

\begin{equation}
\label{loss} - \widehat{C}_{\text{Uno}}(T, \eta) \ = - \ \frac{\sum_{i,k} \Delta_i \, (\hat{G}_{n}^{L}( \tilde{T}_i))^{-2} \, \mathrm{I}
\left( \tilde{T}_i < \tilde{T}_k \right) \mathrm{I} \left(
{\eta}_i > {\eta}_k \right)}{\sum_{i,k} \Delta_i \, (\hat{G}_{n}^{L}( \tilde{T}_i))^{-2} \, \mathrm{I}
\left( \tilde{T}_i < \tilde{T}_k \right) } \ .
\end{equation}

Setting $\mathcal{R} =  - \widehat{C}_{\text{Uno}} (T,\eta)$, however, is unfeasible because $\widehat{C}_{\text{Uno}}(T, \eta)$ is not differentiable with respect to $\eta_i$ and can therefore not be used in a gradient boosting algorithm. To solve this problem, we follow the approach of Ma and Huang \cite{ma2005} and approximate the indicator function in (\ref{loss}) by the sigmoid function $K(u) = 1 / (1 + \exp (-u / \sigma))$. Here, $\sigma$ is a tuning parameter that controls the smoothness of the approximation (details on the choice of $\sigma$ will be given in the Numerical Results section). Replacing the indicator function in (\ref{loss}) by its smoothed version results in the smoothed empirical risk function

\begin{equation}
\label{loss_smooth} - \widehat{C}_{\text{smooth}}(T, \eta) \ = \ - \sum_{i,k} w_{ik} \cdot \frac{1} {1\ +\ \exp \left(\frac{\hat{\eta}_k - \hat{\eta}_i }{\sigma}\right)}
\end{equation}

with weights

\begin{equation}
\label{weights}
w_{ik} \ :=  \ \frac{ \Delta_i \, (\hat{G}_{n}^{L}( \tilde{T}_i))^{-2} \, \mathrm{I}
\left( \tilde{T}_i < \tilde{T}_k \right) }
{ \sum_{i,k} \Delta_i \, (\hat{G}_{n}^{L}( \tilde{T}_i))^{-2} \, \mathrm{I}
\left( \tilde{T}_i < \tilde{T}_k \right)} \ .
\end{equation}

By definition, the smoothed empirical risk $- \widehat{C}_{\text{smooth}}(T, \eta)$ is differentiable with respect to the predictor $\eta_i$. Its derivative is given by

\begin{equation}
\label{grad_smooth} - \frac{\partial \, \widehat{C}_{\text{smooth}}(T, \eta)}{\partial \, \eta_i}\ = - \ {\sum_{k} w_{ik} \, \frac{-\exp \left(\frac{\hat{\eta}_k - \hat{\eta}_i }{\sigma}\right)} {\sigma \, \left(1\ +\ \exp \left(\frac{\hat{\eta}_k - \hat{\eta}_i }{\sigma}\right)\right)}} \ .
\end{equation}

In the next step of the gradient boosting algorithm, the derivative in (\ref{grad_smooth}) is iteratively fitted to a set of base-learners. Typically, an individual base-learner (simple regression tool, e.g., a tree or a simple linear model) is specified for each marker.  To ensure that the estimate of the optimal predictor $\eta^*$ is a {\em linear} combination of the components of $X$, we will apply simple linear models as base-learners (cf. \cite{buehlmann_high}). In other words, each base-learner is a simple linear model with one component of $X$ as input variable. Consequently, there are $p$ base-learners, which will be denoted by $b_l$, $l=1, \dots, p$. Each base-learner refers to one component of $X$ and therefore to one marker (or gene).

The component-wise gradient boosting algorithm for the optimization of the smoothed $C$-index is then given as follows:\\

\line(1,0){460}

\begin{enumerate}
\item[{(1)}] \textbf{Initialize} the estimate of the marker combination $\hat{\eta}^{[0]}$ with offset values. For example, set $\hat{\eta}^{[0]} = \mathbf{0}$, leading to $\hat{\beta}_l^{[0]}=0$ for all components $l=1,\dots,p$. Choose a sufficiently large maximum number of iterations $m_{\text{stop}}$ and set the iteration counter $m$ to 1.

\item[{(2)}] \textbf{Compute} the negative gradient vector by using formula (\ref{grad_smooth}) and evaluate it at the marker combination $\hat{\eta}^{[m-1]}$ of the previous iteration:
\begin{equation*}
U^{[m]} = \left(U_{i}^{[m]}\right)_{i=1,\dots,n} := \left(  \frac{\partial \, \widehat{C}_{\text{smooth}}(T, \hat{\eta}^{[m-1]})}{\partial \, \eta_i}
 \right)_{i=1,\dots,n} \ .
\end{equation*}

\item[{(3)}] \textbf{Fit} the negative gradient vector $U^{[m]}$ separately to each of the components of $X$ via the base-learners $b_l(\cdot)$:
   $$ U^{[m]} \  \stackrel{\rm fitted\ by}{\xrightarrow{\hspace*{1.25cm}}} \
\hat{b}_l^{[m]}(x_l)  \quad \text{for } l=1,...,p.$$

\item[{(4)}] \textbf{Select} the component $l^*$ that best fits the negative gradient vector
according to the least squares criterion, i.e., select the base-learner $b_{l^*}$ defined by

    $$ l^* = \underset{1 \leq l \leq p}{\operatorname{argmin}}\sum_{i=1}^n \left(U_{i}^{[m]} - \hat{b}_{l}^{[m]}(x_l)\right)^2 \ . $$

\item[{(5)}] \textbf{Update} the marker combination $\hat{\eta}$ for this component:

$$ \hat{\eta}^{[m]} \leftarrow \hat{\eta}^{[m-1]} + \text{sl}
\cdot \hat{b}_{l^*}^{[m]}(x_{l^*}) \ , $$

where sl is a small step length $(0 < \text{sl} \ll 1)$. For example, if $\text{sl} = 0.1$, only 10\% of the fit of the base-learner is added to the current marker. This procedure shrinks the effect estimates towards zero, effectively increasing the numerical stability of the update step \cite{friedman_bias, harrell3}.

As only the base learner $\hat{b}_{l^*}$ was selected, only the effect of component $l^*$ is updated ($\hat{\beta}_{l^*}^{[m]} =  \hat{\beta}_{l^*}^{[m-1]} + \text{sl} \cdot \hat{b}_{l^*}^{[m]}(x_{l^*})$) while all other effects stay constant ($\hat{\beta}_{l}^{[m]}  = \hat{\beta}_{l}^{[m-1]}$ for $l \ne l^* $).

\item[{(6)}] \textbf{Stop} if $m = m_{\text{stop}}$. Else increase $m$ by one and go back to step (2).
\end{enumerate}

\line(1,0){460}

\vspace{1cm}

By construction, the proposed boosting algorithm automatically estimates the optimal linear biomarker combination that maximizes the smoothed $C$-index. The principle behind the proposed algorithm is to minimize the empirical risk $\mathcal{R} = -  \widehat{C}_{\text{smooth}}(T, \eta)$  by using gradient descent in function space, where the function space is spanned by the base-learners $b_l$, $l=1,...,p$. In other words, the algorithm iteratively descents the empirical risk  by updating $\hat{\eta}^{[m]}$ via the best fitting base-learner. Because the base-learners are simple linear models (each containing only one possible biomarker as predictor variable) and because the update in step (5) of the algorithm is additive, the final solution $\hat{\eta}^{[m_{\text{stop}}]}$ effectively becomes a linear combination of these markers.

The two main tuning parameters of gradient boosting algorithms are the stopping iteration $m_{\text{stop}}$ and the step length sl. In the literature it has been argued that the choice of the step length is of minor importance for the performance of boosting algorithms \cite{Schmid:Hothorn:boosting-p-Splines}. Generally, a larger step length leads to faster convergence of the algorithm. However, it also increases the risk of overshooting near the minimum of $\mathcal{R}$. In the following sections we will use a fixed step-length of $\text{sl}=0.1$, which is a common recommendation in the literature on gradient boosting \cite{Schmid:Hothorn:boosting-p-Splines, mayr2012importance} (and which is also the default value in the \texttt{R} package  \texttt{mboost} \cite{pkg:mboost:CRAN}). The stopping iteration $m_{\text{stop}}$ is considered to be the most important tuning parameter of boosting algorithms \cite{mayr2012importance}. The optimal value of $m_{\text{stop}}$ is usually determined by using cross-validation techniques \cite{BuhlmannHothorn06}. Small values of $m_{\text{stop}}$  reduce the complexity of the resulting linear combination $\hat{\eta}^{[m_{\text{stop}}]}$ and avoid overfitting via shrinking the effect estimates. In case of boosting the $C$-index, however,  overfitting is less problematic as the predictive performance of $\eta$ is not related to the actual size of the coefficients but to the concordance of the \emph{rankings} between marker values and the observed survival times. As a result, the stopping iteration $m_{\text{stop}}$ in this specific case is less relevant and can be also specified by a fixed large value (e.g., $m_{\text{stop}} = 50000$).

Regarding the boosting algorithm for the smoothed $C$-index, an additional tuning parameter is given by the smoothing parameter $\sigma$. While too large values of $\sigma$ will lead to a poor approximation of the indicator functions in (\ref{loss}), too small values of $\sigma$ might overfit the data (and might therefore result in a decreased prediction accuracy). Details on how to best choose the value of $\sigma$ will be given in the Numerical Results section.

The boosting algorithm presented above is implemented in the add-on software package \texttt{mboost} of the open source statistical programming environment \texttt{R} \cite{Rcite}. The specification of the new {\tt Cindex()} family and a short description of how to apply the algorithm in practice are given in the Appendix.

\subsection*{Evaluation}

After having applied the $C$-index to select the most influential genes (Task 1), and after having used the proposed boosting algorithm to combine the selected genes (Task 2), a final challenge is to evaluate the prediction accuracy of the resulting gene combination (Task 3). Since the $C$-index was used for Tasks 1 and 2, it is also a natural criterion to evaluate the derived marker combination in Task 3. As argued before, it is advantageous from both a methodological perspective as well as from a practical one to use the same criterion for estimation and evaluation of a biomarker combination.

To avoid over-optimistic estimates of prediction accuracy, it is crucial to use different observations for the optimization and  evaluation of the marker signature \cite{harrell3, optimism}. This can be achieved either by using two completely different data sets (\emph{external evaluation}) or by splitting one data-set into a learning sample $\{ (\tilde{T}_i^L, \Delta_i^L, X_i^L),\, i=1,\dots,n\}$ and a test sample $\{ (\tilde{T}_j, \Delta_j, X_j),\, j=1,\dots,N\}$. The learning sample is used to optimize the marker combination while the ``external'' test sample serves only for evaluation purposes. A more elaborate strategy is {\it subsampling} (such as five-fold or ten-fold cross-validation), which is based on multiple random splits of the data. In our numerical analysis, we will use subsampling techniques in combination with stratification to divide the underlying data sets into learning and test samples (for details, see the next section).

When it comes to the $C$-index, two additional points have to be taken into consideration: First, as the task is to obtain the most precise estimation for the discriminatory power, it is no longer necessary to use the smoothed version $\widehat{C}_{\text{smooth}}$ (which was included for numerical reasons in the boosting algorithm) for evaluation. Consequently, we propose to apply the original estimator $\widehat{C}_{\text{Uno}}$ for evaluating biomarker combinations in Task 3. Second, when applying the estimator $\widehat{C}_{\text{Uno}}$ to the observations in a test sample, a natural question is how to calculate the Kaplan-Meier estimator $\hat{G}_{n}^{L}(t)$ of the unconditional survival function of $T_{\text{cens}}$. In principle, there are three possibilities for the calculation of $\hat{G}_{n}^{L}(t)$: The Kaplan-Meier estimator can be computed from either the test or from the training data, or, alternatively, from the combined data set containing all observations in the learning and test samples. Following the principle that all estimation steps should be carried out prior to Task 3, we will base computation of the Kaplan-Meier estimator on the learning data.

\section*{Numerical results}
\subsection*{Simulation study}

We first investigated the performance of our approach based on simulated data. The aim of our simulation study was:

\begin{itemize}
\item[(i)] To analyze if the proposed framework is able to select a small amount of informative markers from a much larger set of candidate variables.
\item[(ii)] To check if gradient boosting is able to derive the optimal combination $\eta$ of the selected markers,  and to compare its performance to competing Cox-based estimation schemes.
\item[(iii)] To investigate the effect of the smoothing parameter $\sigma$ that controls the smoothness inside the sigmoid function, as well as potential effects of the sample size and the censoring rate on the performance of our approach.
\end{itemize}

The simulated survival times are generated via a log-logistic distribution for accelerated failure time (AFT) models \cite{kleinmoesch}. They are based on the model equation $\log(T) = \mu + \phi W $, where $T$ is the survival time, $\mu$ and $\phi$ are location and scale parameters, and $W$ is a noise variable, following a standard logistic distribution.  As a result, the density function for realizations $t_i$ can be written as

\begin{equation}
f_{\text{dens}}(t_i|\mu_i,\phi_i) = \frac{\exp\left((t_i-\mu_i)/\phi_i\right)}{\phi_i\left(1 + \exp\left((t_i-\mu_i)/\phi_i \right)\right)^2}
\end{equation}

with $\mathbb{E}(T) =\mu $ and $\text{Var} (T) = \frac{\pi^2}{3\phi^2}$. The $p=1000$ possible markers $X_1,....,X_{1000}$ were drawn from a multivariate normal distribution with pairwise correlation ($\rho=0.5$). The \textit{true} underlying combinations of the predictors were given by

\begin{eqnarray*}
         \mu_i &=& \eta_{\mu}    = 1.5 + 1.5x_1 + x_2 - x_3 - 1.5x_4  \ , \\
          \log(\phi_i) &=& \eta_{\phi} = -1 + 2 x_1 -2 x_2 +x_3 -x_4 \ .
\end{eqnarray*}

Note that only four out of $1000$ possible markers have an actual effect on the survival time. Furthermore, those four markers do not only contribute to the location parameter $\mu$ but also to the scale parameter $\phi$ (cf. \cite{gamboostlss:2012}) -- a setting where standard survival analysis clearly becomes problematic. Additionally to the survival times $T$, we generated an independent sample of censoring times $T_{\text{cens}}$ and defined the observed survival time by $\tilde{T} := \min(T, T_{\text{cens}})$ leading to independent censoring of 50\% of the observations.

\begin{figure}[t]
\begin{center}
\includegraphics[width=\textwidth]{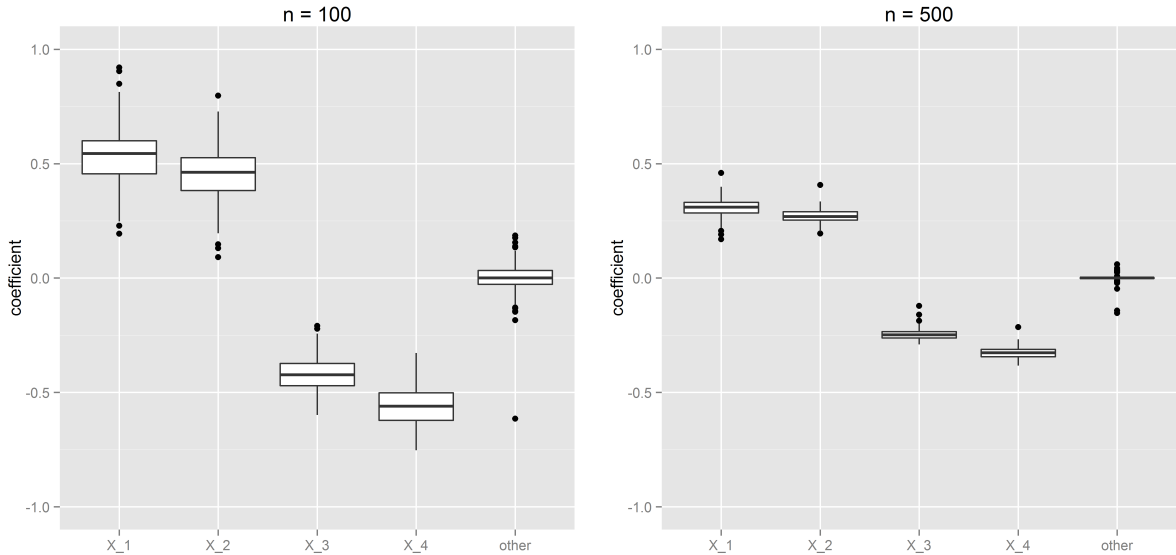}
\end{center}
\caption{
{\bf Coefficient estimates for $p^*=5$ pre-selected markers obtained from 100 simulation runs}. The marker combinations were optimized via gradient boosting based on training samples of size $n=100$ (left) and $n=500$ (right).  Boxplots represent the empirical distribution of the resulting coefficients. Only markers $X_1$ to $X_4$ had an actual effect on the survival time.}
\label{Box_coefs}
\end{figure}

In order to answer the above questions, we investigated the performance of our framework in all three tasks that are necessary to develop new gene signatures in practice (Tasks 1--3 described in the Introduction). To avoid over-optimism and biased results, we simulated separate data sets for the different tasks.  In $B=100$ simulation runs, we first simulated a data set  $\{(\tilde{T}_i, \Delta_i, X_i),\, i=1,\dots,1000\}$ with $1000$ observations to pre-select the most influential predictors based on the $C$-index (Task 1). The optimal combination $\eta$ of those predictors was later estimated (Task 2) by our boosting algorithm based on smaller training samples $\{ (\tilde{T}_i^L, \Delta_i^L, X_i^L),\, i=1,\dots,n\}$ of size $n$.  For the final external evaluation of the prediction accuracy (Task 3) we simulated a separate test data set $\{(\tilde{T}_j, \Delta_j, X_j),\, j=1,\dots,N\}$  with $N= 1000$.

For Task 1, we first pre-selected a subset of $p^*$ predictors from the $p=1000$ available markers. We ranked the predictors based on their individual values of $\hat{C}_{\text{Uno}}$ and included only the $p^* = \{5, 10, 30\}$ best-performing markers in the boosting algorithm. The results suggest that the $C$-index is clearly able to identify markers that are truly related to the outcome: Although all predictors had a relatively high pairwise correlation ($\rho=0.5$), the four informative markers had a selection probability of 98.5\% for $p^*=5$ (99\% for $p^*=10$ and 99.5\% for $p^*=30$).

\rowcolors{2}{gray!25}{white}
\begin{table}[t]
\centering
\begin{tabular}{|ccc|cccc|}
  \hline
 \rowcolor{white}
 \multicolumn{3}{|c|}{setting}  &  \multicolumn{4}{c|}{method} \\
 \rowcolor{white}
 $n$& $p^*$ & $cens.$& $C$-index boosting & Cox lasso & Cox ridge &  \textit{true} $C$-index\\
  \hline
  100 &  5  &  50\%    & 0.764 (0.04) & 0.731 (0.06)& 0.739 (0.04)& 0.779 \\
  100 &  10  &  50\%    & 0.746 (0.06) & 0.709 (0.08)& 0.707 (0.06)& 0.779  \\
  100 &  30  &  50\%    & 0.689 (0.07) & 0.673 (0.11)& 0.637 (0.07)& 0.779 \\
  100 &  5  &  30\%    & 0.820 (0.02) & 0.774 (0.04)& 0.724 (0.04)& 0.830 \\
  100 &  5  &  70\%    & 0.668 (0.10) & 0.628 (0.10)& 0.593 (0.11)& 0.690 \\
   50 &  5  &  50\%    & 0.741 (0.07)& 0.662 (0.18)& 0.722 (0.09)& 0.772  \\
   200 &  5  &  50\%    & 0.774 (0.02)& 0.748 (0.04)& 0.752 (0.04)& 0.782  \\
   500 &  5  &  50\%    & 0.778 (0.03)& 0.759 (0.03)& 0.760 (0.02)& 0.781  \\
   \hline
\end{tabular}
\caption{
\textbf{Results of the simulation study.} Comparison of the discriminatory power resulting from boosting the $C$-index with competing approaches. Numbers refer to the median value and interquartile range (in parentheses) of the final $\hat{C}_{\text{Uno}}$ on 100 simulation runs. The \textit{true} $C$-index refers to the discriminatory power resulting from the true combination of predictors with known coefficients. The amount of pre-selected genes is denoted as $p^*$, $n$ is the size of the training samples and \emph{cens.} refers to the censoring rate.} \label{tab:sims}
\end{table}

To find the optimal combination $\eta$ of the pre-selected markers (Task 2), we applied the proposed boosting approach on training samples with size $n=100$. The resulting coefficients for $p^* =5$ and smoothing parameter $\sigma = 0.1$ are presented in Figure~\ref{Box_coefs}. The boosting algorithm seems to be able to derive the optimal combination of the pre-selected markers, as the structure displayed by the coefficients is essentially the same as the one of the underlying \textit{true} combination $\eta_{\mu}$. The discriminatory power of the resulting biomarker does not depend on the absolute size of the coefficients: As the $C$-index is based solely on the concordance between biomarker and survival time, what matters in practice is the \emph{relative} size of the coefficients. As can be seen from Figure~\ref{Box_coefs}, the estimated positive effect for $x_1$ is larger than the one for $x_2$. On the other hand, the negative effect of $x_4$  is correctly estimated to be more pronounced than the the one of $x_3$. The coefficient of the falsely selected marker is on average close to zero.

\begin{figure}[t]
\begin{center}
\includegraphics[width=\textwidth]{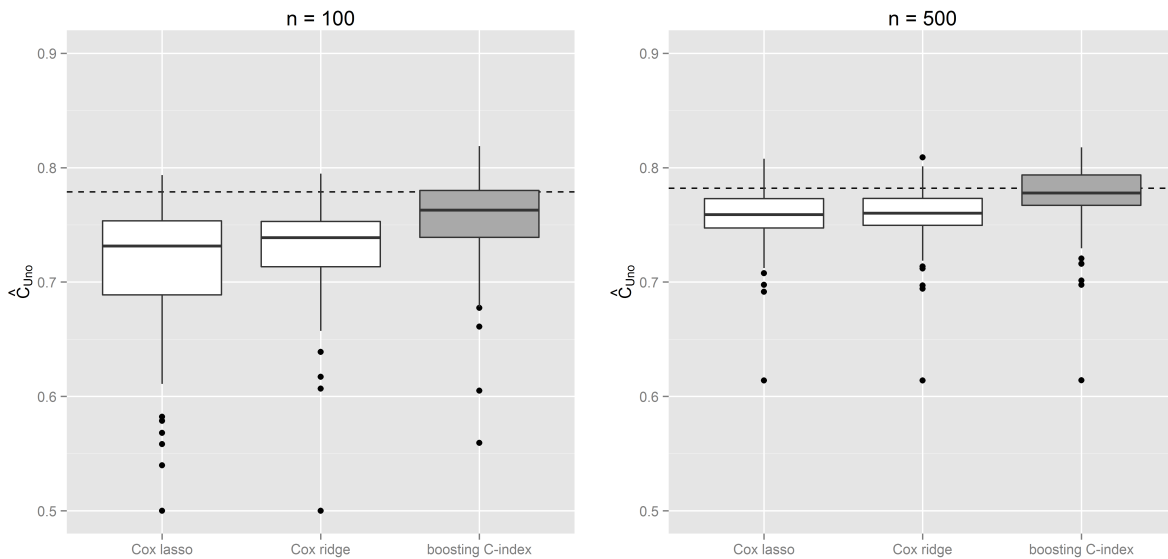}
\end{center}
\caption{
{\bf Simulation results for the discriminatory power obtained via the proposed $C$-index boosting approach and competing Cox-based estimation schemes.} The marker combinations were optimized via the different approaches based on training samples of size $n=100$ (left) and $n=500$ (right).  Boxplots represent the empirical distribution of the resulting $\hat{C}_{\text{Uno}}$ on corresponding test samples.  The dotted line corresponds to the discriminatory power resulting from the \emph{true} combination of predictors with known coefficients.}
\label{Box_sims}
\end{figure}

In a third step, we evaluated the performance of the resulting optimized marker combinations (Task 3) on separate test samples. The resulting estimates $\hat{C}_{\text{Uno}}$ for different simulation settings are presented in Table~\ref{tab:sims}.  The highest discriminatory power (median $\hat{C}_{\text{Uno}} = 0.763$, range = 0.559--0.819) can be observed for $p^* = 5$, which is closest to the true number of informative markers. We compared the performance of our proposed algorithm to penalized Cox regression approaches such as Cox-Lasso \cite{tibshCoxlasso, goemanlasso} and Cox regression with ridge-penalization \cite{li2002Ridge, gui2005penalized} -- see Figure~\ref{Box_sims}. The proposed boosting approach clearly outperforms the competing estimation schemes, supporting our view that applying traditional Cox regression might be suboptimal if the discriminatory power is the performance criterion of interest. We additionally computed the optimal $C$-index resulting from the \emph{true} marker combination $\eta_{\mu}$ with known coefficients. The values of the true $C$-index on the test samples are on average only slightly better than the ones of boosting the concordance index (median $\hat{C}_{\text{Uno}} = 0.778$  -- see Table~\ref{tab:sims}).

To evaluate the possible effects of different sample sizes and censoring rates we modified the mean censoring time leading to approximate censoring rates of 30\% and 70\% and generated training samples of size $n=\{50, 200, 500\}$. Results are included in Table~\ref{tab:sims}. As expected, the $C$-index resulting from our framework increases as censoring rates become small (median $\hat{C}_{\text{Uno}} = 0.820$, range = 0.736--0.858)  and decreases in settings with a large proportion of censored observations (median $\hat{C}_{\text{Uno}} = 0.668$, range = 0.421--0.776). However, the same effect can be observed for the true $C$-index resulting from the true marker combination $\eta_{\mu}$ (30\% censoring $\hat{C}_{\text{Uno}} = 0.830$, 70\% censoring $\hat{C}_{\text{Uno}} = 0.690$). For larger training samples, the variance of the coefficient estimates decreases (see Figure~\ref{Box_coefs}). As a result, the discriminatory power resulting from our boosting algorithm improves (for $n=500$, median $\hat{C}_{\text{Uno}} = 0.778$, range = 0.614--0.818) and gets nearly as good as the true $C$-index ($\hat{C}_{\text{Uno}} = 0.781$). This finding further confirms the ability of our approach to find the most optimal marker combination possible -- see Figure~\ref{Box_sims}. Note that also the true $C$-index differs slightly between the different sample sizes, as the training sample enters in $\hat{C}_{\text{Uno}}$ via the Kaplan-Meier estimator $\hat{G}_{n}^{L}(t)$.

To investigate the effect of the smoothing parameter inside the sigmoid function, we additionally applied our boosting procedure for every simulation setting with different values of $\sigma$.  The resulting estimates $\hat{C}_{\text{Uno}}$  are presented in Table~\ref{tab:simssigma}. Compared to the effects of the sample size or the number of pre-selected markers $p^*$, the smoothing parameter $\sigma$ only seems to have a minor effect on the performance of our algorithm. In light of these empirical results, we recommend to apply a fixed small value (e.g.,  $\sigma = 0.1$, which is also the default value in the \texttt{Cindex()} family for the \texttt{mboost} package \cite{pkg:mboost:CRAN} -- see the Appendix).

\rowcolors{2}{gray!25}{white}
\begin{table}[t]
\centering
\begin{tabular}{|ccc|ccccc|}
  \hline
   \rowcolor{white}
   \multicolumn{3}{|c|}{setting}  &  \multicolumn{5}{c|}{smoothing parameter} \\
    \rowcolor{white}
  $n$& $p^*$ & $cens.$ & $\sigma=0.5$ & $\sigma=0.25$ & $\sigma=0.1$ & $\sigma=0.075$ & $\sigma=0.05$ \\
  \hline
   100 &  5  &  50\%    & 0.738 (0.06) & 0.757 (0.05)& 0.764 (0.04)& 0.763 (0.04)& 0.762 (0.04) \\
   100 &  10  &  50\%    & 0.728 (0.06) & 0.744 (0.06)& 0.746 (0.06)& 0.746 (0.06)& 0.741 (0.05) \\
  100 &  30  &  50\%    & 0.700 (0.06) & 0.702 (0.07)& 0.689 (0.07)& 0.683 (0.07)& 0.666 (0.07)\\
 100 &  5  &  30\%    & 0.802 (0.03) & 0.815 (0.02)& 0.820 (0.02)& 0.821 (0.02)& 0.822 (0.02)\\
 100 &  5  &  70\%    &  0.665 (0.10) & 0.667 (0.10)& 0.668 (0.10)& 0.665 (0.10)& 0.661 (0.10)\\
  50 &  5  &  50\%    & 0.719 (0.07)& 0.737 (0.07)& 0.741 (0.07)& 0.740 (0.06)& 0.725 (0.06)\\
  200 &  5  &  50\%    & 0.743 (0.05)& 0.768 (0.03)& 0.774 (0.02)& 0.775 (0.02)& 0.778 (0.02)\\
  500 &  5  &  50\%    & 0.723 (0.05)& 0.769 (0.02)& 0.778 (0.03)& 0.779 (0.03)& 0.781 (0.03)\\
   \hline
\end{tabular}
\caption{\textbf{Simulation results for different values of the smoothing parameter.} Comparison of the discriminatory power resulting from the gradient boosting approach when applying different values of the smoothing parameter $\sigma$. Numbers refer to to the median value and interquartile range (in parentheses) of the final $\hat{C}_{\text{Uno}}$ on 100 simulation runs.  The amount of pre-selected genes is denoted as $p^*$, $n$ is the size of the training samples and \emph{cens.} refers to the censoring rate. We recommend to use the value $\sigma = 0.1$, which is also the default value of the new \texttt{Cindex} family for the R add-on package \texttt{mboost}.}
\label{tab:simssigma}
\end{table}

For both approaches to fit penalized Cox regression (Cox lasso, Cox ridge), we applied the \texttt{R} add-on package \texttt{penalized} \cite{penalized}. In order to evaluate $\hat{C}_{\text{Uno}}$, we used the \texttt{UnoC()} function implemented in the \texttt{survAUC} package \cite{survM}.

\subsection*{Applications to predict the time to distant metastases}

In the next step, we further analyzed the performance of our gradient boosting algorithm in two applications to estimate and evaluate the optimal combination of pre-selected biomarkers. All markers are used to predict the time to distant metastases in breast cancer patients. As in the simulation study, we compared the results of our proposed algorithm to Cox regression with lasso and ridge penalization. Additionally, we considered four competing boosting approaches for survival analysis, which do not directly optimize the $C$-index. The first is classical Cox regression, estimated via gradient boosting, while the other three are parametric accelerated failure-time (AFT) models assuming a Weibull, log-normal or log-logistic distribution \cite{hothorn2006survival, Schmid:Hothorn:AFT-boost}. For all boosting approaches (Weibull AFT boosting, loglog-AFT boosting and Cox boosting) we used the corresponding pre-implemented functions of the \texttt{mboost} package. To ensure comparability,  we used the same linear base-learners as described above for all boosting approaches.

To ensure that the combined predictor did not only work on the data it was derived from but also on ``external'' validation data, we carried out a subsampling procedure for both data sets to generate $B=100$ different learning samples $\{ (\tilde{T}_i^L, \Delta_i^L, X_i^L),\, i=1,\dots,n\}$ and test samples $\{(\tilde{T}_j,\Delta_j,X_j), j=1,\dots,N \}$, respectively. Consequently, we randomly split the corresponding data sets to use $2/3$ of the observations as learning sample in order to optimize the biomarker combination $\hat{\eta}$. To ensure an equal distribution of patients with and without an observed development of distant metastases we applied stratified sampling.  Correspondingly, the $1/3$ of the observations not included in the learning sample were used to evaluate the resulting predictions via the $C$-index. As a result,  for every data set and every method we computed 100 different combinations $\eta$ and 100 corresponding values of $\hat{C}_{\text{Uno}}$.

\subsubsection*{Breast cancer data by Desmedt et al.}

Desmedt et al. \cite{desmedt} collected a data set of 196 node-negative breast cancer patients to validate a 76-gene expression signature developed by Wang et al. \cite{wanglancet}. The signature, which is based on Affymetrix microarrays, was developed separately for estrogen-receptor (ER) positive patients (60 genes) and ER-negative patients (16 genes). In addition to the expression levels of the 76 genes, four clinical predictor variables were considered (tumor size, estrogen receptor (ER)  status, grade of the tumor and patient age). The data are publicly available on GEO (\texttt{http://www.ncbi.nlm.nih.gov/geo}, accession number GSE 7390).

Similar to Wang et al. \cite{wanglancet}, we used the time from diagnosis to distant metastases as primary outcome and considered the 76 genes together with the four clinical predictors. Observed metastasis-free survival ranged from 125 days to 3652 days, with 79.08\% of the survival times being censored.

\begin{figure}[t]
\begin{center}
\includegraphics[width=\textwidth]{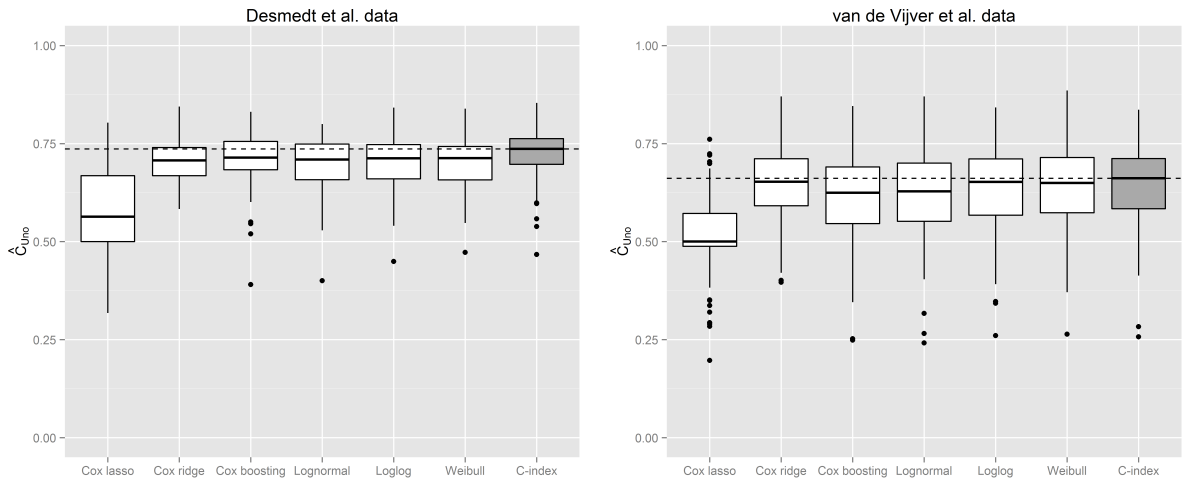}
\end{center}
\caption{
{\bf Comparing the discriminatory power of biomarker combinations to predict the time to distant metastases resulting from} the proposed $C$-index boosting approach with competing estimation schemes. The plot on the left refers to the Desmedt et al. data, whereas the plot on the right presents results from the van de Vijver et al. data. All biomarker combinations were optimized via the corresponding algorithms based on the same 100 learning samples.  Boxplots represent the empirical distribution of the resulting $\hat{C}_{\text{Uno}}$ on corresponding test samples. The dotted line corresponds to the median $C$-index resulting from the new approach. }
\label{Box_data}
\end{figure}

The main results of our analysis are presented in Figure~\ref{Box_data}. As expected, the unified framework to estimate and evaluate the optimal marker signature based on the $C$-index is not only methodologically more consistent than the Cox and AFT approaches, but also leads to to marker signatures that show a higher discriminatory power on external or future data (median $\hat{C}_{\text{Uno}} = 0.736$, range = 0.467--0.854). As discussed in the methodological section, it is crucial to evaluate the discriminatory power on external data: the estimated $C$-index on the training sample was more than 35\% higher (median $\hat{C}_{\text{Uno}} = 0.986$) and hence extremely over-optimistic \cite{harrell3, optimism}.



Considering the interpretation  of the resulting coefficient estimates for the clinical predictors, it is crucial to note that boosting methods for the $C$-index and the AFT models yield biomarker combinations $\eta^*$ where larger values indicate \textit{longer} predicted survival times. On the other hand, classical Cox regression models rely on the hazard; higher values are hence associated with \textit{smaller} survival times. If this is taken into account, results from the different approaches were in fact very similar.  Both age of the patients and size of the tumor had a negative effect on the time to recurrence for all seven approaches. The same holds true for the tumor grade \textit{poor differentiation} which resulted in a negative effect compared to \textit{good differentiation} and \textit{intermediate differentiation}. A positive ER status, on the other hand, was associated with a larger metastasis-free survival in all approaches. Regarding the coefficients of the 76 genes, results from our approach to boost the $C$-index were highly correlated to the ones of the other four boosting approaches (which rely on the same base-learners) --  absolute correlation coefficients computed from the 100 subsamples ranged from 0.77 to 0.99. Also coefficients resulting from the popular ridge-penalized Cox regression were highly correlated with the ones from our approach -- absolute correlation coefficients ranged from 0.47 to 0.84.

\subsubsection*{Breast cancer data by van de Vijver et al.}

The second data set consists of $144$ lymph node positive breast cancer patients that was collected by the Netherlands Cancer Institute \cite{vandevijver}. The data set, which is publicly available as part of the \texttt{R} add-on package \texttt{penalized} \cite{penalized}, was used by van de Vijver et al. \cite{vandevijver} to validate a 70-gene signature for metastasis-free survival after surgery developed by van't Veer et al. \cite{vantveer}. In addition to the expression levels of the 70 genes, the data set contains five clinical predictor variables (tumor diameter, number of affected lymph nodes, ER status, grade of the tumor and patient age). Observed metastasis-free survival times ranged from $0.055$~months to $17.660$~months, with $67\%$ of the survival times being censored.

Resulting values of the $C$-index of the new approach and the six considered competitors are presented in Figure~\ref{Box_data}. The improvement from applying the proposed unified framework compared to boosting the Cox proportional hazard model or applying ridge-penalized Cox regression was much less pronounced than in the previous data set. However, on average, boosting the $C$-index still led to the best combination of markers regarding the discriminatory power (median $\hat{C}_{\text{Uno}} = 0.662$, range = 0.257--0.836). Interestingly, as in the previous data set, the lasso penalized Cox regression was clearly outperformed by the ridge-penalized competitor (which has been suggested for this specific data set by van Houwelingen et al. \cite{Houwelingen}). Furthermore, the ridge-penalized approach performed at least as good as the considered boosting approaches (except the new approach to boost the $C$-index). As in the previous data set, we again additionally evaluated the $C$-index on the training sample in order to assess the resulting over-optimism. As expected, the estimated $C$-index on the training sample was extremely biased (median $\hat{C}_{\text{Uno}} = 0.973$).


The resulting coefficients for the clinical predictors were again comparable for the seven different approaches. A positive ER status was associated with a larger metastasis-free survival for all seven approaches, the same also holds true for the age of the patient. On the other hand, the size of the tumor, the number of affected lymph nodes and a poor tumor grade resulted for all approaches in a negative effect on the survival time. Regarding the coefficients of the  79 genes, the highest correlation could again be observed for the boosting algorithms: Absolute correlation coefficients obtained from the 100 subsamples ranged from 0.64 to 0.95. Correlation between coefficients resulting from our approach to boost the $C$-index and the ones from ridge-penalized Cox regression was slightly lower, it ranging from 0.30 to 0.82.

\section*{Discussion}

In this article we have proposed a framework for the development of survival prediction rules that is based on the concordance index for time-to-event data ($C$-index). As the $C$-index is an easy-to-interpret measure of the accuracy of survival predictions (based on clinical or molecular data), it has become an important tool in medical decision making. Generally, the focus of the $C$-index is on measuring the ``discriminatory power'' of a prediction rule: It quantifies how well the rankings of the survival times and the values of a biomarker (or marker combinations) in a sample agree. In particular, the $C$-index is methodologically different from measures that evaluate how well a prediction rule is ``calibrated'' (i.e., from measures that quantify ``how closely the predicted probabilities agree numerically with the actual outcomes'' \cite{pencina}).
Specifically, prediction rules that are well calibrated do not necessarily have a high discriminatory power (and vice versa). 

While several authors have proposed the use of the $C$-index for feature selection \cite{masong} and the evaluation of molecular signatures \cite{zhang_plosone, zhao_plosone}, the main contribution of this paper is a new approach for the {\em derivation of marker combinations} that is based directly on the $C$-index. Consequently, when using the proposed method, it is no longer necessary to rely on traditional methods such as Cox regression -- which focus on the derivation of well-calibrated prediction rules instead of well-disciminating prediction rules and may therefore be suboptimal when the optimization of the discriminatory power is of main interest.


Conceptually, our approach is in direct line with recent articles by Ma and Huang \cite{ma2005}, Wang and Chang \cite{wang2011marker} and Schmid et al. \cite{schmidetal:sagm} who developed a set of algorithms for the optimization of discrimination measures for binary outcomes (such as the area under the curve (AUC) and the partial area under the curve and (PAUC)). Because the $C$-index is in fact a summary measure of a correspondingly defined AUC measure for time-to-event data \cite{heagerty05}, our optimization technique relies on similar methodological concepts, such as the application of boosting methods and the use of smoothed indicator functions.

A possible future extension of our approach might be to include the task of selecting the most influential genes in the proposed boosting algorithm. While our simulation study and the breast-cancer examples were based on the pre-selection of genes, the proposed boosting method could also be applied directly to high-dimensional molecular data, so that Tasks 1 and 2 are effectively combined. This can be accomplished by optimizing the stopping iteration so that only a (low-dimensional) subset of the candidate genes is incorporated in the resulting marker combination (``early stopping'', cf. \cite{mayr2012importance}). Further research is warranted on the issues of early stopping and automated feature selection in the case of boosting the concordance index for survival data.

The results of our empirical analysis suggest that the new approach is competitive with state-of-the-art methods for the derivation of marker combinations. As demonstrated in the Numerical Results section, the resulting marker combinations are not only easy to compute and have a meaningful interpretation but can also lead to a higher discriminatory power of the resulting gene signatures.

\section*{Acknowledgments}

The authors thank Sergej Potapov for his help with the analysis of the breast cancer data. The work of Matthias Schmid and Andreas Mayr was supported by Deutsche Forschungsgemeinschaft (DFG) (www.dfg.de), grant SCHM 2966/1-1. The funders had no role in study design, data collection and analysis, decision to publish, or preparation of the manuscript.

\small
\bibliography{literature_v2}

\normalsize
\clearpage
\section*{Appendix}

\subsection*{Gradient boosting of the $C$-index}

The concept of boosting was first introduced in the field of machine learning \cite{Schapire89, Freund90}. The basic idea is to boost the accuracy of a relatively weak performing classificator (termed ``base-learner'') to a more accurate prediction via iteratively applying the base-learner and aggregating its solutions. Generally, the concept of boosting leads to a drastically improved prediction accuracy compared to a single solution of the base-learner \cite{Ridgeway99}.  This basic concept was later adapted to fit statistical regression models in a forward stagewise fashion \cite{friedmanetal2000, buehlmann_high}. One of the main advantages of this approach is the interpretability of the final solution, which is basically the same as in any other statistical model \cite{BuhlmannHothorn06}. This can not be achieved with competing machine learning algorithms as Support Vector Machines \cite{vapnik} or Random Forests \cite{rf}. Specifically, the boosting approach can be used to develop prediction rules for survival outcomes \cite{hothorn2006survival, Schmid:Hothorn:AFT-boost, binder2009_survival}. Although there exist also likelihood-based approaches for boosting \cite{TutzBinder}, we will focus here on gradient-based boosting \cite{BuhlmannHothorn06} as it is the better fitting approach for boosting the distribution-free $C$-index.

The most flexible implementation of gradient boosting is the \texttt{mboost} \cite{pkg:mboost:CRAN} add-on package for the Open Source programming environment \texttt{R} \cite{Rcite}. The \texttt{mboost} package contains a large variety of different pre-implemented base-learners and loss functions, that can be combined by the user via different fitting functions. For a tutorial on the how to apply the package for practical data analysis, see \cite{hofner2012model}.

To apply gradient boosting to optimize linear biomarker combinations w.r.t. the $C$-index in the version of Uno et al. \cite{unoC}, it is necessary to specify the newly developed \texttt{Cindex()} family inside the \texttt{glmboost()} function.

The \texttt{Cindex} family object includes the sigmoid function $K(u) = 1 / (1 + \exp (-u / \sigma))$ as approximation of the indicator functions in the estimated $C$-index. The sigmoid function is evaluated inside the {\tt R} functions \texttt{approxGrad()} and \texttt{approxLoss()}, which are part of the \texttt{Cindex} object. The weights

\begin{equation}
w_{ik} \ :=  \ \frac{ \Delta_i \, (\hat{G}_{n}^{L}( \tilde{T}_i))^{-2} \, \mathrm{I}
\left( \tilde{T}_i < \tilde{T}_k \right) }
{ \sum_{i,k} \Delta_i \, (\hat{G}_{n}^{L}( \tilde{T}_i))^{-2} \, \mathrm{I}
\left( \tilde{T}_i < \tilde{T}_k \right)} \
\end{equation}

are computed via the internal function \texttt{compute\_weights()} for both the empirical risk

\begin{equation}
 - \widehat{C}_{\text{smooth}}(T, \eta) \ = \ - \sum_{i,k} w_{ik} \cdot \frac{1} {1\ +\ \exp \left(\frac{\hat{\eta}_k - \hat{\eta}_i }{\sigma}\right)} \
\end{equation}

(implemented in the \texttt{risk()} function) as well as for the negative gradient

\begin{equation}
 - \frac{\partial \, \widehat{C}_{\text{smooth}}(T, \eta)}{\partial \, \eta_i}\ = - \ {\sum_{k} w_{ik} \, \frac{-\exp \left(\frac{\hat{\eta}_k - \hat{\eta}_i }{\sigma}\right)} {\sigma \, \left(1\ +\ \exp \left(\frac{\hat{\eta}_k - \hat{\eta}_i }{\sigma}\right)\right)}} \
\end{equation}

(implemented in the \texttt{ngradient()} function).

Those different functions that define the optimization problem are finally plugged into the \texttt{mboost} specific \texttt{Family()} function to build a new {\tt boost$\_$family}. Details on how to implement user-specific families in \texttt{mboost} are presented in the Appendix of \cite{hofner2012model}. The complete \texttt{Cindex} object is then given as follows:

\begin{small}
\begin{verbatim}
Cindex <- function (sigma = 0.1) {

      approxGrad <- function(x) {                     ## sigmoid function for gradient
        exp(x/sigma) / (sigma * (1 + exp(x/sigma))^2)
      }
      approxLoss <- function(x) {                     ## sigmoid function for loss
        1 / (1 + exp(x / sigma))
      }

      compute_weights <- function(y, w = 1){          ## compute weights
        ipcw_wow <- IPCweights(y[w != 0,])
        ipcw <- numeric(nrow(y))
        ipcw[w!=0] <- ipcw_wow
        survtime <- y[,1]
        n <- nrow(y)
        wweights <- matrix( (ipcw)^2, nrow = n, ncol = n)
        weightsj <- matrix(survtime, nrow = n, ncol = n)
        weightsk <- matrix(survtime, nrow = n, ncol = n, byrow = TRUE)
        weightsI <- (weightsj < weightsk) + 0
        wweights <- wweights * weightsI
        Wmat <- w %o% w
        wweights <- wweights * Wmat
        wweights <- wweights / sum(wweights)
        rm(weightsI); rm(weightsk); rm(weightsj)
      return(wweights)
      }

      ngradient = function(y, f, w = 1) {              ## negative gradient
        if (!all(w %in% c(0,1)))
          stop(sQuote("weights"), " must be either 0 or 1 for family ",
               sQuote("UnoC"))
        survtime <- y[,1]
        event <- y[,2]
        if (length(w) == 1) w <- rep(1, length(event))
        if (length(f) == 1) {
          f <- rep(f, length(survtime))
        }
        n <- length(survtime)
        etaj <- matrix(f, nrow = n, ncol = n, byrow = TRUE)
        etak <- matrix(f, nrow = n, ncol = n)
        etaMat <- etak - etaj
        rm(etaj); rm(etak);
        weights_out <- compute_weights(y, w)
        M1 <- approxGrad(etaMat) * weights_out
        ng <- colSums(M1) - rowSums(M1)
        return(ng)
      }

      risk = function(y, f, w = 1) {                   ## empirical risk
        survtime <- y[,1]
        event <- y[,2]
        if (length(f) == 1) {
          f <- rep(f, length(y))
        }
        n <- length(survtime)

        etaj <- matrix(f, nrow = n, ncol = n, byrow = TRUE)
        etak <- matrix(f, nrow = n, ncol = n)
        etaMat <- (etak - etaj)
        rm(etaj); rm(etak);
        weights_out <- compute_weights(y, w)
        M1 <- approxLoss(etaMat) * weights_out
        return(- sum(M1))
      }

        Family(                                        ## build the family object
          ngradient = ngradient,
          risk = risk,
          weights = "zeroone",
          offset = function(y, w = 1) {0},
          check_y = function(y) {
            if (!inherits(y, "Surv"))
              stop("response is not an object of class ", sQuote("Surv"),
                   " but ", sQuote("family = UnoC()"))
            y},
          rclass = function(f){},
          name = paste("Concordance Probability by Uno")
          )
}


\end{verbatim}
\end{small}

\subsection*{Application}

We will briefly demonstrate how to apply the \texttt{Cindex} family in practice to derive the optimal combination of pre-selected biomarkers. We will use the van de Vijver et al. \cite{vandevijver} data set of $144$ lymph node positive breast cancer patients that was also considered in the main article. The data set is publicly available as part of the \texttt{R} add-on package \texttt{penalized} \cite{penalized}. The 70-gene signature for metastasis-free survival after surgery was originally developed by van't Veer et al. \cite{vantveer}.

We first split the data set in 100 training observations and 44 test observations. To ensure better readability of the code we do not carry out stratified subsampling but just use the first 100 patients as training sample. Model fitting is carried out by the \texttt{glmboost()} function of the \texttt{mboost} package.  As linear models are the default base-learners for \texttt{glmboost()}, no additional base-learner has to be specified. As appropriate \texttt{family} object we specify the \texttt{Cindex} family described above.

For evaluating the discriminatory power of the resulting prediction on test data, we use the \texttt{UnoC()} function of the \texttt{survAUC} package \cite{survM}. It implements the unbiased estimator  $\widehat{C}_{\text{Uno}}$, as proposed by Uno et al. \cite{unoC}.

\begin{small}
\begin{verbatim}
## load add-on packages
library(penalized) ## for the data set
library(mboost)    ## for boosting
library(survAUC)   ## for evaluation

data(nki70)        ## loading the data

source("Cindex.R") ##  loading the family defined above

## split the data set in training and test sample (simplified):
dtrain <- nki70[1:100,]
dtest  <- nki70[101:144,]

## fit a model via the glmboost() function
## formula : defines the candidate model; the response is the survival
##           object Surv(time, event); via '~ ." all remaining variables
##           in the data set serve as possible predictors
## family  : defines the optimization problem (in this case the C-index)
##           sigma is the smoothing parameter of the sigmoid function that
##           approximates the indicator functions. The default value is 0.1.
## control : defines other boosting-specific tuning parameters like the
##           stopping iteration mstop or the step-length nu; trace = TRUE is
##           only for convenience (shows the trace of the empirical risk).
## data    : defines the data set -> training sample

mod1 <- glmboost(Surv(time,  event) ~ .,  family = Cindex(sigma = 0.1),
                 control = boost_control(mstop = 500, trace = TRUE, nu = 0.1),
                 data = dtrain)

## The stopping iteration can be changes via simple indexing:
mod1 <- mod1[50000] ## Long runtime: 50000 iterations
                    ## takes at least a couple of minutes on a standard machine

## Now take a look at the resulting combination
coef(mod1)

## Prediction on test data
preds <-  predict(mod1, newdata = dtest)

## Evaluate the discriminatory power
UnoC(Surv(dtrain$time, dtrain$event),  Surv(dtest$time, dtest$event), lpnew = -preds)
\end{verbatim}
\end{small}
\end{document}